\documentclass{article}

\usepackage{microtype}
\usepackage{graphicx}
\usepackage{subfigure}
\usepackage{booktabs} % for professional tables
\usepackage[accepted]{icml2025}

\usepackage{hyperref}

\usepackage{amsmath}
\usepackage{amssymb}
\usepackage{mathtools}
\usepackage{amsthm}

\usepackage[capitalize,noabbrev]{cleveref}

\theoremstyle{plain}

\theoremstyle{definition}

\theoremstyle{remark}

\usepackage[textsize=tiny]{todonotes}

\icmltitlerunning{Financial Sentiment Analysis Using FinBERT with Application in Predicting Stock Movement}

\begin{document}

\twocolumn[
\icmltitle{Financial Sentiment Analysis Using FinBERT with Application in Predicting Stock Movement}



\icmlsetsymbol{equal}{*}

\begin{icmlauthorlist}
% \icmlauthor{Firstname1 Lastname1}{equal,yyy}
% \icmlauthor{Firstname2 Lastname2}{equal,yyy,comp}
% \icmlauthor{Firstname3 Lastname3}{comp}
\icmlauthor{Qingyun Zeng}{sch1}
\icmlauthor{Tingsong Jiang}{sch2}
\end{icmlauthorlist}

\icmlaffiliation{sch1}{University of Pennslyvania Philadelphia, United states}
\icmlaffiliation{sch2}{University of Rochester Rochester, United states}

\icmlcorrespondingauthor{Qingyun Zeng}{qze@sas.upenn.edu}
% \icmlcorrespondingauthor{Tingsong Jiang}{jiangtingsong1992@gmail.com}

\icmlkeywords{Machine Learning, ICML}

\vskip 0.3in
]



\printAffiliationsAndNotice{\icmlEqualContribution} % otherwise use the standard text.

\begin{abstract}
In this study, we integrate sentiment analysis within a financial framework by leveraging FinBERT, a fine-tuned BERT model specialized for financial text, to construct an advanced deep learning model based on Long Short-Term Memory (LSTM) networks. Our objective is to forecast financial market trends with greater accuracy. To evaluate our model's predictive capabilities, we apply it to a comprehensive dataset of stock market news and perform a comparative analysis against standard BERT, standalone LSTM, and the traditional ARIMA models. Our findings indicate that incorporating sentiment analysis significantly enhances the model's ability to anticipate market fluctuations. Furthermore, we propose a suite of optimization techniques aimed at refining the model's performance, paving the way for more robust and reliable market prediction tools in the field of AI-driven finance.
\end{abstract}

\section{Introduction}
The advent of Transformer-based models like BERT (Bidirectional Encoder Representations from Transformers) \cite{bert19} revolutionized natural language processing (NLP), offering powerful contextual representations that proved highly successful across diverse tasks. There has been various variants of BERT which focu on different tasks in NLP and different domain of language models. FinBERT, which is introduced in \cite{finbert19}, is the first contextual pretrained language models which is trained specifically on a large scale of financial communication corpora. This is originally designed for the purpose of financial sentiment analysis especially focus on the need of capital market practitioners and researchers. There has been various applications of FinBERT in research, for example, \cite{finbert20} studies financial text mining, \cite{finbert20b} studies financial communications, \cite{finbert22} studies information extraction in finance etc. 

Our main task in this paper is to use sentiment-based information in predicting the movement of the financial market. Financial data is usually presented in time series data, for example, stock prices. Traditional statistical method such as ARIMA has been used widely in the modelling of financial markets. However, statistical models like ARIMA does not capture well the nonlinearity in the financial data.
People moved to deep learning methods especially in LSTM in studying financial time series, for example in \cite{lstm22}. 
Our basic hypothesis is that the financial sentiment by news, social media, etc. have strong correlations to the movement of financial markets (stock market, foreign exchange, bond, etc.). This hypothesis is based on Efficient Market Hypothesis (EMH), which states that share prices reflect all information. Hence we are able to predict financial market using all those information. In particular, we believe that news contain a large amount of information. This hypothesis is a well-known, but the how to build quantitative models and making the prediction is still a challenging problem. We propose a model which takes advantage of Transformer based contextual language model for analyzing the sentiment in financial markets, and then using LSTM-based deep learning model in the prediction of the exact movement of the financial market by the sentiment. One thing need to be pointed our that, we are not doing casual analysis here, as market could influence news and news might also influence the market. We just study the relation of these two objects, and quantify this correlation. 

We find the the LSTM model in general perform much better than ARIMA model, and sentiment analysis using FinBERT out perform LSTM, whereras FinBERT combined with LSTM acheived the best performance. This result complies with our assumption and anticipation.

\section{Related Work}

{\it Transformer-based attention network for stock move-
ment prediction}\cite{trans22} proposed a novel model Transformer Encoder-based Attention Network (TEANet) framework for the prediction of stock price movement, and apply it to a trading simulation to test the validity of this model. Their model does not use sentiment analysis, and uses the transformer model for the time series data modelling, which is different than our sentiment-LSTM structure. However, their model would be helpful in our future work that we can replace the LSTM model by the TEANet for the sequencing modelling.

{\it Bert-based financial sen-
timent index and lstm-based stock return predictability}\cite{bertlstm} studied similar problem as us, but with the base BERT for their sentiment analysis part. They proposed {\it Financial Sentiment Index}, which incorporated both the contextual channel and two other type of information sources which are option-implied sentiment based on the  risk-neutral implied skewness, and the market implied sentiment based on market data. These three channel are then put together for the prediction of stock return. Our sentiment analysis is just using the raw sentiment information from the FinBERT, which might be tuned to incorporate Financial Sentiment Index in the future work. Their prediction model is also based on LSTM, which is then compared to a more traditional econometric model called VAR (Vector Autoregression). The VAR required lots of assumptions, while the noval model does not require. For the results, LSTM performed much better than the VAR which implied that LSTM captured nonlinearlity of the market information. Our model uses FinBERT in the sentiment analysis part, which is supposed to have much better performance in financial data sets. Hence we anticipate our model would our perform using base BERT.

Traditionally, the ARIMA model is used in modelling financial time series. This method does not capture well the nonlinearity of the information. {\it Attention-based cnn-lstm and xgboost hybrid model for stock prediction}\cite{att2022} proposed a mixed model which combines the attention-based-covolutional neural network, LTSM, and then fine-tune by XGBoost. This paper did not use sentiment analysis. First, the model uses the traditional ARIMA for preprocessing the data. Then it uses attention-based CNN to extract deep features of the original stock data. After that, LSTM is used to mine the long term time series features. Finally, XGBoost is used to fine-tune the model performance. It turned out that attention-based CNN with Bi-LSTM achieved the the best performance across all base line models as well as single CNN or LSTM or BiLSTM. The model is this paper is similar to the sequencing model part of ours but with more stacking of various different models. Our model will not use the fusion model in the times series part, but in the future work we might adapt their ideas and method in improving the performance.

\section{Data}
Our study utilizes a combination of pre-trained models, financial news archives, and stock market data, sourced as follows:

\begin{itemize}
    \item {\bf Pre-trained Language Models:} We employ {\it base BERT uncased} and {\it FinBERT} (`ProsusAI/finbert`) obtained from the Hugging Face Model Hub. FinBERT, specifically, is chosen for its specialization in financial text.
    \item {\bf Historical Financial News Archive (Primary Textual Data):} This dataset, publicly available on Kaggle.com, comprises news archives for over 800 US companies, spanning 12 years up to 2019. It includes crucial features like date, stock ticker, news title, and body. Its prior use in sentiment analysis studies makes it a suitable foundation for our research. This dataset is joined with market data to facilitate both sentiment analysis and market movement prediction.
    \item {\bf Daily Financial News for 6000+ Stocks (Supplementary Textual Data):} Also sourced from Kaggle.com, this dataset offers a broader range of news items, albeit with less detailed information per entry (e.g., primarily headers). It serves to complement our primary news archive.
    \item {\bf Stock Market Data:} We retrieve historical stock market data (e.g., open price, close price, percentage change) for the tickers identified in our news datasets using the yfinance API. This numerical data is a critical input for our LSTM model, alongside the derived sentiment scores.
\end{itemize}

For our primary analysis, we focus on the {\it Historical financial news archive}. We augment this by merging the {\tt analyst\_ratings\_processed.csv} file from the {\it Daily Financial News for 6000+ Stocks} dataset, after removing irrelevant columns. The combined dataset contains 1,056,471 records, each featuring news titles, publication dates, and stock tickers. It's important to note that the news data extends up to 2019, a factor considered in our analysis and model training timeframe.

\begin{figure}
    {\includegraphics[width=0.5\textwidth]{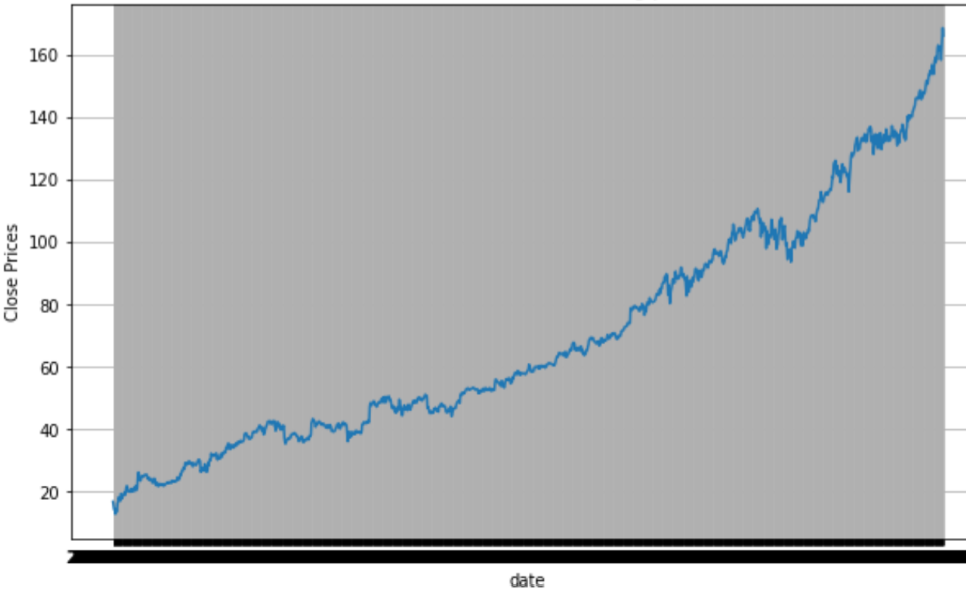}}
    \caption{A plot of MSFT closing price}
\end{figure}

\section{Data Preprocessing and Feature Engineering}

Our data preprocessing pipeline involves several key steps to ensure the quality and relevance of the input data for modeling:

\textbf{Merging datasets:} We combined two financial news datasets—the Historical Financial News Archive and the Daily Financial News for 6000+ Stocks (specifically, the \texttt{analyst\_ratings\_processed.csv} file from the latter after removing irrelevant columns, and the main news data after dropping metadata columns such as category, provider, and URL to focus on titles, dates, and tickers). These were concatenated, and duplicates were removed based on unique news titles. We then merged this textual data with numerical stock market data (open, high, low, close prices, and trading volume) obtained using the yfinance API.

\textbf{Sentiment Extraction:} We utilized the pre-trained FinBERT model (`ProsusAI/finbert`) from the Hugging Face library. Each news title was tokenized and passed through FinBERT to obtain sentiment probabilities for positive, negative, and neutral classes. These probabilities were averaged daily per stock ticker, resulting in a daily aggregated sentiment score computed as:
\[
\text{SentimentScore} = P(\text{positive}) - P(\text{negative})
\]

\textbf{Numerical Sentiment Index (NSI):} To quantify sentiment directly from market price movements, we introduce the Numerical Sentiment Index (NSI), defined as:
\[
\text{NSI}_t =
\begin{cases}
    1 & \text{if } \text{return}_t > s \\
    0 & \text{if } -s \leq \text{return}_t \leq s \\
    -1 & \text{if } \text{return}_t < -s
\end{cases}
\]
where the daily return is calculated as:
\[
\text{return}_t = \frac{\text{ClosePrice}_t - \text{OpenPrice}_t}{\text{OpenPrice}_t}
\]
with a threshold $s=0.01$. The NSI provides a market-based sentiment label useful for fine-tuning FinBERT or analyzing correlations between textual sentiment and market movements.

\textbf{Handling Missing Values and Non-Trading Days:} We handled missing values by dropping incomplete records. Additionally, we accounted for non-trading days (weekends and U.S. public holidays) by aligning news sentiment scores with the market data of the next available trading day, ensuring temporal consistency between sentiment and price movements. This alignment was performed programmatically by identifying subsequent business days.

\textbf{Normalization and Scaling:} Financial time series data, such as stock prices (open, high, low, close) and trading volume, often exhibit varying scales and magnitudes. To address this, we applied Min-Max scaling to these numerical features, transforming them into a [0, 1] range. A Min-Max scaler was fitted on the training data for the input features (open, high, low, volume). A separate Min-Max scaler was fitted on the training data for the target variable (closing price); this allows for the model's predictions to be inverse-transformed back to their original price scale for evaluation. For each feature $x$, the scaled value $x_{\text{scaled}}$ is computed as:
\[
x_{\text{scaled}} = \frac{x - x_{\text{min}}}{x_{\text{max}} - x_{\text{min}}}
\]
where $x_{\text{min}}$ and $x_{\text{max}}$ are the minimum and maximum values of the feature observed in the training dataset. This normalization step is crucial, particularly for neural network models like LSTMs, as it helps to stabilize the training process, accelerate convergence, and prevent features with inherently larger numerical values from disproportionately influencing model learning.

\textbf{Temporal Train-Test Split:} To evaluate our model's forecasting ability in a realistic setting, we employed a strict temporal train-test split. The dataset was ordered chronologically, with the initial 90\% of the data used for training and the subsequent 10\% reserved for testing. This chronological partitioning is paramount in time series forecasting to prevent data leakage, where information from the future (test set) inadvertently influences the model during training, which can occur with random splits. Random splits are generally unsuitable for time-series data as they disrupt the temporal dependencies. Our choice of a 90/10 split was also informed by initial experiments with an 80/20 split, which yielded less satisfactory results, potentially due to the increased difficulty of predicting longer-term market dynamics influenced by significant, unforeseen events.

This comprehensive preprocessing and feature engineering pipeline ensures robust and meaningful inputs for our subsequent modeling steps, improving the reliability and interpretability of our predictive results.

\section{Sentiment Analysis Methodology}

Our sentiment analysis methodology leverages the pre-trained FinBERT model (`ProsusAI/finbert`) from the Hugging Face Transformers library. FinBERT is a transformer-based language model specifically fine-tuned on financial text corpora, including Reuters TRC2 and Financial PhraseBank datasets \cite{finbank14}, making it particularly suitable for financial sentiment analysis tasks.

\textbf{Tokenization and Encoding:} Each news title is tokenized using the FinBERT tokenizer, which converts textual data into numerical representations suitable for transformer-based models. Specifically, we apply padding and truncation to ensure uniform input lengths, with a maximum sequence length of 512 tokens to accommodate the transformer model's input constraints.

\textbf{Sentiment Probability Extraction:} Tokenized inputs are passed through the FinBERT model, which outputs logits corresponding to three sentiment classes: positive, negative, and neutral. We apply a softmax function to these logits to obtain sentiment probabilities:
\[
P(\text{class}_i) = \frac{e^{\text{logit}_i}}{\sum_{j} e^{\text{logit}_j}}, \quad i \in \{\text{positive}, \text{negative}, \text{neutral}\}
\]

\textbf{Daily Aggregation of Sentiment Scores:} To align sentiment data with daily stock market data, we aggregate sentiment probabilities across all news titles published on the same day for each stock ticker. Specifically, we compute the daily aggregated sentiment score as:
\[
\text{SentimentScore}_t = \frac{1}{N_t}\sum_{n=1}^{N_t}\left[P_n(\text{positive}) - P_n(\text{negative})\right]
\]
where $N_t$ is the number of news articles for a given stock ticker on day $t$, and $P_n(\text{positive})$, $P_n(\text{negative})$ are the sentiment probabilities for the $n$-th news article.

\textbf{Numerical Sentiment Index (NSI):} Additionally, we utilize the previously defined NSI (see Section 4) to provide market-based sentiment labels. These labels can be leveraged for further fine-tuning of FinBERT or for analyzing correlations between textual sentiment and market movements.

This structured sentiment analysis methodology ensures that we effectively capture and quantify the sentiment embedded in financial news, providing robust inputs for our predictive modeling framework.
\section{The Model}

Our proposed model architecture integrates financial sentiment analysis with a time-series forecasting module. It is designed to predict future stock price movements by leveraging both textual news data and historical market data. The model consists of two primary components:

\begin{enumerate}
    \item {\bf Financial Sentiment Analysis Module}:
    This module utilizes the pre-trained FinBERT model (`ProsusAI/finbert`) from the Hugging Face Model Hub. FinBERT is a BERT-based model specifically fine-tuned on a large corpus of financial text, including the Reuters TRC2 dataset and the Financial PhraseBank \cite{finbank14}, making it adept at understanding financial nomenclature and sentiment.
    For each news item, we input the news title (and optionally the body, if available and deemed beneficial after experimentation) into FinBERT. The model outputs probabilities for three sentiment classes: positive, negative, and neutral. We convert these probabilities into a single numerical sentiment score. A common approach is to calculate this as: $\text{SentimentScore} = P(\text{positive}) - P(\text{negative})$. This continuous score, ranging from -1 to 1, quantifies the sentiment intensity and direction. No further fine-tuning of the FinBERT model was performed in this specific study to maintain a focus on its off-the-shelf capabilities combined with the LSTM. The extracted sentiment score for each news item, aggregated per day per stock ticker, serves as a key input feature for the subsequent LSTM module.

    \item {\bf LSTM-based Time-Series Forecasting Module}:
    This module is responsible for predicting stock price movements using the sentiment scores and historical stock market data.
\begin{itemize}
        \item {\bf Input Features}: For each time step $t$ (representing a trading day), the input vector to the LSTM, $X_t$, is constructed by concatenating the daily aggregated sentiment score (from the Financial Sentiment Analysis Module) with relevant historical stock market features. These typically include: lagged values of the closing price, opening price, daily high, daily low, and trading volume. For instance:
        \item {\bf Input Features}: For each time step $t$ (representing a trading day), the input vector to the LSTM, $X_t$, is constructed by concatenating the daily aggregated sentiment score (from the Financial Sentiment Analysis Module) with relevant historical stock market features. These typically include: lagged values of the closing price, opening price, daily high, daily low, and trading volume. For instance:
        \begin{equation*}
            X_t = \begin{multlined}[t]
                    [S_t, C_{t-1}, O_{t-1}, H_{t-1}, L_{t-1}, V_{t-1}, \dots, \\
                    C_{t-k}, O_{t-k}, H_{t-k}, L_{t-k}, V_{t-k}]
                  \end{multlined}
        \end{equation*}
where $S_t$ is the sentiment score for day $t$; $C, O, H, L, V$ are close, open, high, low, and volume respectively; and $k$ is the look-back window size (set to 60 days in our experiments). All numerical features are normalized using Min-Max scaling to a range of [0, 1] before being fed into the LSTM to ensure stable training.
        
        \item {\bf LSTM Architecture}: Our LSTM network is structured as follows:
        \begin{enumerate}
            \item An initial LSTM layer with 100 hidden units. The \texttt{return\_sequences=True} parameter is set, meaning this layer outputs the full sequence of hidden states for each input time step. This is suitable for stacking LSTM layers. Standard \texttt{tanh} activation is used for the recurrent step and \texttt{sigmoid} for gates.
            \item A second LSTM layer, also with 100 hidden units. For this layer, \texttt{return\_sequences=False} is set, so it only outputs the hidden state of the final time step. This aggregates information over the entire input sequence.
            \item A Dense (fully connected) layer with 25 units and a ReLU (Rectified Linear Unit) activation function, $\text{ReLU}(x) = \max(0, x)$. This layer introduces further non-linearity.
            \item A final Dense output layer with a single unit and a linear activation function. This unit predicts the target variable, which is typically the next day's closing price or price change.
        \end{enumerate}
                \item {\bf Training Details}: The LSTM-based forecasting module was trained to minimize the Mean Squared Error (MSE) between the predicted and actual stock closing prices. We employed the Adam optimizer \cite{kingma2014adam}, a widely-used adaptive learning rate optimization algorithm, with an initial learning rate of 0.001. Models were trained for up to 100 epochs with a batch size of 1. A batch size of 1 means that model weights were updated after each sample, which can be beneficial for capturing fine-grained patterns in sequential data, though it can also lead to noisier gradient updates. To prevent overfitting and select the best performing model, we utilized an early stopping mechanism: training was halted if the validation loss (MSE on the validation set) did not improve for 20 consecutive epochs. The validation set comprised 10\% of the training data, selected chronologically from the end of the training portion. This setup was chosen based on common practices and empirical performance in preliminary experiments.

    \end{itemize}
\end{enumerate}

We compare our model performance to the following baseline models:
\begin{enumerate}
    \item {\bf Simple ARIMA model}: A traditional statistical model for time series forecasting. Preprocessing involved checking for stationarity using the Augmented Dickey-Fuller test, applying log transformations to stabilize variance, and differencing to achieve stationarity. The ARIMA(1,1,2) model (selected based on ACF/PACF plots and information criteria) was then fitted to the historical log-transformed price data.

    \item {\bf LSTM on numerical data only}: An LSTM model with the same architecture as described above, but trained solely on historical stock market data without sentiment features.
    \item {\bf FinBERT with DNN}: Sentiment scores from FinBERT were concatenated with numerical stock data (open, high, low, volume, and lagged close prices, all MinMax scaled). This combined feature set was fed into a Deep Neural Network consisting of an input normalization layer, followed by two dense layers of 256 units each with Leaky ReLU activation, and a final dense output layer with a single unit and linear activation. The model was trained using the Adam optimizer (learning rate 1e-3, with decay) for up to 1000 epochs (batch size 100), with early stopping (patience 200) based on validation loss (20\% split).

    \item {\bf Base BERT with LSTM}: An identical architecture to our proposed model, but using sentiment scores derived from a standard pre-trained BERT (e.g., `bert-base-uncased`) instead of FinBERT, to assess the specific contribution of financial domain adaptation.
\end{enumerate}
\begin{figure}
    {\includegraphics[width=0.5\textwidth]{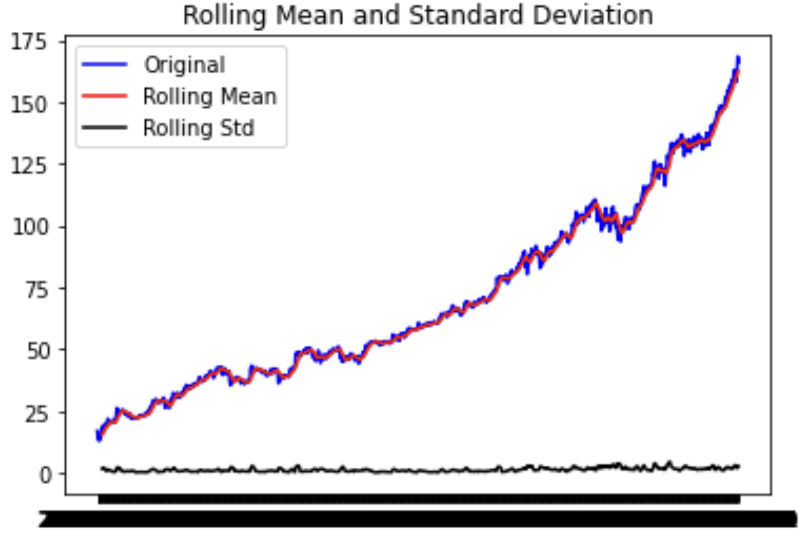}} 
    \caption{A plot of rolling mean and standard deviation for MSFT closing price}
    \label{fig:foobar}
\end{figure}
We use Min Square Error (MSE) as our main metric in both training and validation process since our target variable (close price) is numerical, and our model is a time series model. For ARIMA, we also use Mean Absolute Error (MAE) and Root Mean Squared Error (RMSE) during the tuning process.

\section{Experiments}
\begin{table*}[t]
\caption{Training and Testing loss on different models}
\label{tab:model_performance}
\vskip 0.15in
\begin{center}
\begin{small}
\begin{sc}
\begin{tabular}{lcc}
\toprule
\textbf{Models} & \textbf{Training Loss} & \textbf{Val loss} \\
\midrule
ARIMA & N/A  & 0.00975 \\
LSTM & 1.0206e-03 & 1.1937e-03 \\
BERT + LSTM & 1.0216e-04   & 3.3132e-04\\
FinBERT + DNN & 1.4674e-04 & 3.4177e-04 \\
FinBERT + LSTM (Proposed) & 9.1701e-05 & 3.1975e-04 \\
\bottomrule
\end{tabular}
\end{sc}
\end{small}
\end{center}
\vskip -0.1in
\end{table*}
We started from the sentiment analysis using the FinBERT. Using the HuggingFace API, we are able to get sentiment from each financial news. We utilize the Numerical Sentiment Index (NSI), as defined in Section 4, to quantify market-based sentiment. The NSI is particularly useful for fine-tuning FinBERT and studying the correlation between textual sentiment and price-based sentiment.
    
Our task is to predict close price. In order to avoid the long-run inflation and size of different stocks, we use MinMax scalar before the model, and normalize the training variables. We trained our model with 90-10 split based on date, i.e. the first 90\%  days as training data and the remaining as testing data. The reason is the it is very hard to predict long term behavior of the Financial market, as there are too many factors can infuence the fluctuations of Financial markets, and some of them might be completely unexpected, for example, COVID-19, war, etc. We actually started from 80-20 splits, and the result was not so satisfying, which did not outperform a lot than our baseline LSTM model.

\section{Analysis}
From the result we see that our FinBERT sentiment analysis + LSTM model acheives the best training and validation loss. Base BERT + LSTM follows closely with it, which implies that our model has not fully taken advantage of the FinBERT language model since we only use the raw sentiment scores from the off-the-shelf `ProsusAI/finbert` model. Future work should explore fine-tuning FinBERT on task-specific data (e.g., using the NSI proposed in Section 5) or on a larger, more contemporary financial news corpus to potentially enhance its domain-specific capabilities for this prediction task. In this paper, we only look at the sentiment from the titles of news, which contain of relatively  small amount information and sentiment. This suggest us also trying to see other informations, for example, social network data like twitter, weibo, etc.

We also see that ARIMA actually capture the long run trend well, whereas our model is able to fit the short run trend which corresponding to the movement in short amount of the time pretty good. For a very short training time compared to other deep learning based models, AIRMA achieves MSE: 0.00975, MAE: 0.07154 and RMSE: 0.09876. As we see from the plot of predicted close price our model tend to under predict the close price in the long run. This suggest that in the future work, we might adopt similar method as in \cite{att2022}, which preprocesses the data by ARIMA first, and then feed it into the deep neural network. 

During the training process, we also see the the loss dropped significantly even just over 3 epochs, which is due to small size of the data. Note that even the whole training set is small, we only trained and tested by individual stock, and each stock only contains several thousands data on average. For example, the raw data for MSFT consists of 7875 rows. After the data processing, we have even less with 1440 rows left. This is challeging for deep learning based method since the model tends to overfit, which means that the model might capture more about the noise in the market, rather than the effective information. As we know that in stock market and even in financial markets in general, there has been strong correlations everywhere. We need to find a way to figure out these correlation, so that we might be able to put all stock data together and make a combined model.

\begin{center}
\begin{figure}

\includegraphics[width=3in]{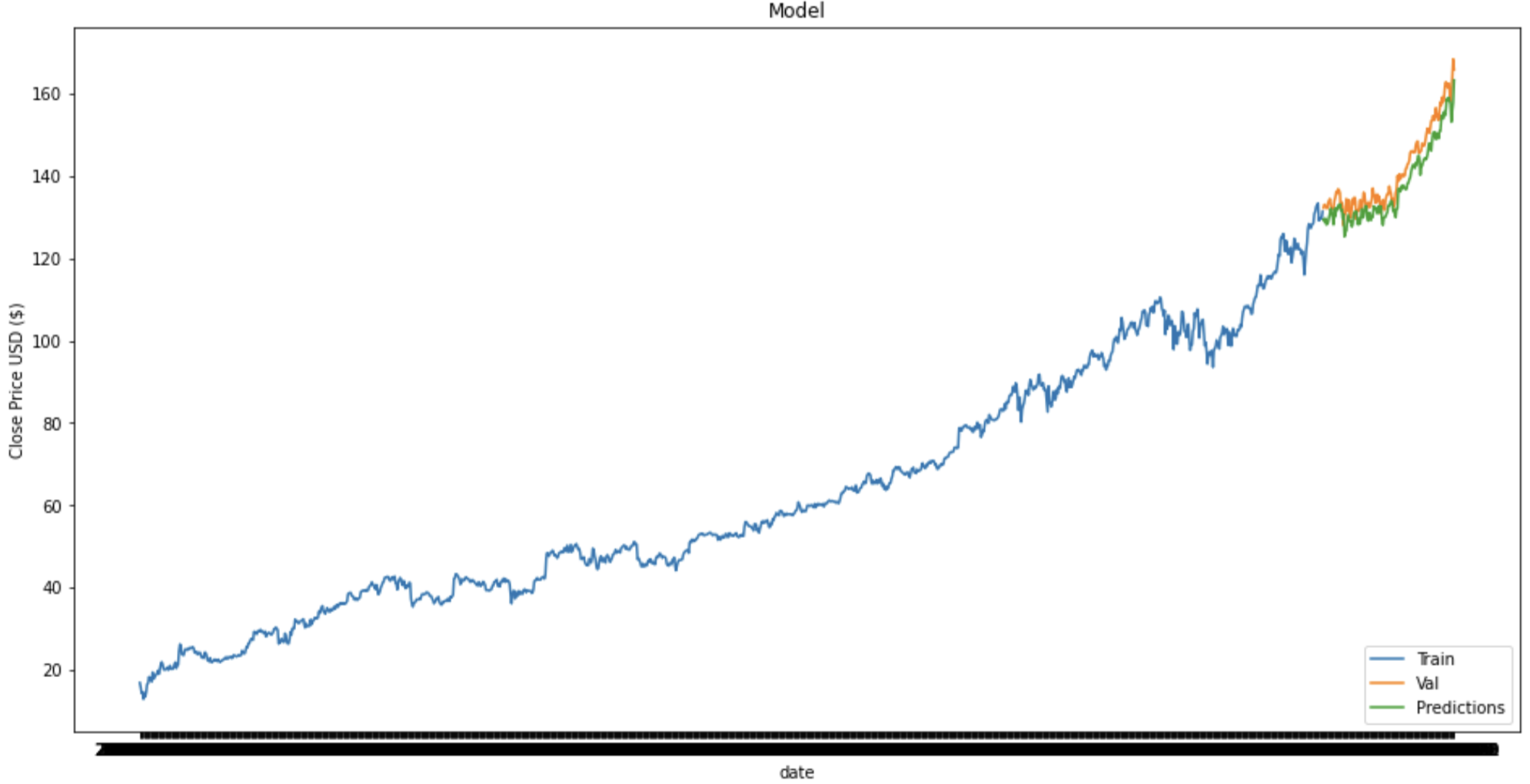}
\caption{\label{fig:price_prediction} % Changed label
Plot of predicted close price and real close price}
\end{figure}
\end{center}
\subsection{Error Analysis and Limitations}
Despite the promising results, our study has several limitations. Firstly, the dataset size for individual stocks, after preprocessing, can be relatively small for training complex deep learning models. For instance, the MSFT dataset reduced to 1440 rows after processing. This limited data availability increases the risk of overfitting, where the model might learn noise specific to the training period rather than generalizable market patterns. This was observed during training where loss dropped significantly within a few epochs.

Secondly, our current sentiment analysis relies solely on news titles. While titles are designed to be informative, they contain a limited amount of textual information compared to the full news article bodies. This might restrict the depth of sentiment that can be extracted.

Thirdly, while FinBERT outperformed base BERT, the improvement margin suggests that the off-the-shelf FinBERT model might not be fully optimized for the specific nuances of our dataset or prediction task.

Future improvements could address these limitations by:
\begin{itemize}
    \item Exploring techniques to augment the dataset or incorporate data from multiple related stocks, leveraging inter-stock correlations, to increase the effective training size.
    \item Incorporating the full text of news articles or sentiment from other sources like social media (e.g., Twitter) to provide richer contextual information.
    \item Investigating further fine-tuning of FinBERT on a larger, more contemporary financial news corpus, or specifically on data labeled with our NSI, to enhance its sentiment extraction capabilities for this task.
\end{itemize}

\section{Conclusion}
In this paper, we presented a model that leverages financial sentiment analysis to enhance the prediction of stock market movements. Our approach combined the specialized FinBERT (`ProsusAI/finbert`), a transformer-based language model trained on financial corpora, with LSTM networks to interpret and utilize time-series data effectively. We demonstrated that this fusion outperforms traditional models such as ARIMA, as well as LSTM models without sentiment analysis, in capturing the nuances of market fluctuations.

Despite the encouraging results, our study identified several areas for improvement and further research. Key directions for future work include:
\begin{enumerate}
    \item \textbf{Hybrid Modeling with ARIMA:} The model exhibits a tendency to underpredict long-term trends. Future work could explore integrating ARIMA for preprocessing the time series data to capture long-range dependencies, before feeding the residuals or ARIMA predictions as features into the LSTM module, similar to hybrid approaches suggested by \cite{att2022}.
    \item \textbf{Data Augmentation and Richer Sentiment Sources:} Limitations in the dataset size for individual stocks led to concerns about overfitting. Future work should explore the inclusion of a broader array of financial news, the full body of news articles, or sentiment data from social media platforms to enrich the training set and provide more comprehensive sentiment signals.
    \item \textbf{Advanced FinBERT Fine-tuning:} While utilizing pre-trained FinBERT showed advantages, the margin of improvement over base BERT was modest. This highlights an opportunity for future work. Building upon findings such as those in "Is Domain Adaptation Worth Your Investment? Comparing BERT and FinBERT on Financial Tasks" \cite{domain21}, further fine-tuning of FinBERT on a larger, more contemporary financial news corpus, or specifically on a dataset labeled with our Numerical Sentiment Index (NSI), could significantly amplify the benefits of domain-specific modeling for this prediction task. Preliminary explorations with alternative labeling strategies for fine-tuning underscored the importance of aligning the fine-tuning task closely with the downstream prediction objective, reinforcing the potential of NSI-based fine-tuning.
    \item \textbf{Exploring Advanced Time-Series Architectures:} While LSTM networks are effective, newer transformer-based architectures designed for time-series forecasting, such as TEANet \cite{trans22}, could be investigated as an alternative to or in conjunction with the LSTM module to potentially capture more complex temporal patterns.
\end{enumerate}

In conclusion, our sentiment-based stock market prediction model represents a promising step forward in the application of AI to financial forecasting. The interplay between news sentiment and market behavior, while complex, can be harnessed to produce more accurate and reliable predictions. As we continue to refine our model by exploring these avenues, we anticipate further improvements in performance and a deeper understanding of the intricate dynamics at play in financial markets.

\bibliography{Finbert}

\begin{thebibliography}{12}
\providecommand{\natexlab}[1]{#1}
\providecommand{\url}[1]{\texttt{#1}}
\expandafter\ifx\csname urlstyle\endcsname\relax
  \providecommand{\doi}[1]{doi: #1}\else
  \providecommand{\doi}{doi: \begingroup \urlstyle{rm}\Url}\fi

\bibitem[Araci(2019)]{finbert19}
Araci, D.
\newblock Finbert: Financial sentiment analysis with pre-trained language models.
\newblock \emph{ArXiv}, abs/1908.10063, 2019.

\bibitem[Devlin et~al.(2019)Devlin, Chang, Lee, and Toutanova]{bert19}
Devlin, J., Chang, M.-W., Lee, K., and Toutanova, K.
\newblock {BERT}: Pre-training of deep bidirectional transformers for language understanding.
\newblock In \emph{Proceedings of the 2019 Conference of the North {A}merican Chapter of the Association for Computational Linguistics: Human Language Technologies, Volume 1 (Long and Short Papers)}, pp.\  4171--4186, Minneapolis, Minnesota, June 2019. Association for Computational Linguistics.
\newblock \doi{10.18653/v1/N19-1423}.
\newblock URL \url{https://aclanthology.org/N19-1423}.

\bibitem[Fjellstrom(2022)]{lstm22}
Fjellstrom, C.
\newblock Long short-term memory neural network for financial time series.
\newblock pp.\  3496--3504, 12 2022.
\newblock \doi{10.1109/BigData55660.2022.10020784}.

\bibitem[Hiew et~al.(2019)Hiew, Huang, Mou, Li, Wu, and Xu]{bertlstm}
Hiew, J. Z.-G., Huang, X., Mou, H., Li, D., Wu, Q., and Xu, Y.
\newblock Bert-based financial sentiment index and lstm-based stock return predictability.
\newblock \emph{arXiv: Statistical Finance}, 2019.

\bibitem[Huang et~al.(2023)Huang, Wang, and Yang]{finbert22}
Huang, A., Wang, H., and Yang, Y.
\newblock {F}in{BERT}: A large language model for extracting information from financial text.
\newblock \emph{Contemporary Accounting Research}, 40\penalty0 (2):\penalty0 1311--1344, 2023.
\newblock \doi{10.1111/1911-3846.12835}.

\bibitem[Kingma \& Ba(2014)Kingma and Ba]{kingma2014adam}
Kingma, D.~P. and Ba, J.
\newblock Adam: A method for stochastic optimization.
\newblock \emph{arXiv preprint arXiv:1412.6980}, 2014.

\bibitem[Liu et~al.(2020)Liu, Huang, Huang, Li, and Zhao]{finbert20}
Liu, Z., Huang, D., Huang, K., Li, Z., and Zhao, J.
\newblock {F}in{BERT}: A pre-trained financial language representation model for financial text mining.
\newblock In \emph{Proceedings of the Twenty-Ninth International Joint Conference on Artificial Intelligence, {IJCAI-20}}, pp.\  4513--4519. ijcai.org, 2020.
\newblock \doi{10.24963/ijcai.2020/623}.

\bibitem[Malo et~al.(2014)Malo, Sinha, Takala, Korhonen, and Wallenius]{finbank14}
Malo, P., Sinha, A., Takala, P., Korhonen, P., and Wallenius, J.
\newblock Good debt or bad debt: Detecting semantic orientations in economic texts.
\newblock \emph{Journal of the American Society for Information Science and Technology}, 04 2014.
\newblock \doi{10.1002/asi.23062}.

\bibitem[Peng et~al.(2021)Peng, Chersoni, Hsu, and Huang]{domain21}
Peng, B., Chersoni, E., Hsu, Y.-Y., and Huang, C.-R.
\newblock Is domain adaptation worth your investment? comparing bert and finbert on financial tasks.
\newblock pp.\  37--44, 01 2021.
\newblock \doi{10.18653/v1/2021.econlp-1.5}.

\bibitem[Shi et~al.(2022)Shi, Hu, Mo, and Wu]{att2022}
Shi, Z., Hu, Y., Mo, G., and Wu, J.
\newblock Attention-based cnn-lstm and xgboost hybrid model for stock prediction.
\newblock \emph{ArXiv}, abs/2204.02623, 2022.

\bibitem[Y~Yang et~al.(2020)Y~Yang, Uy, and Huang]{finbert20b}
Y~Yang, K.Z.~Zhang, P.~K., Uy, M. C.~S., and Huang, A.
\newblock Finbert: A pretrained language model for financial communications.
\newblock \emph{ArXiv}, abs/2006.08097, 2020.

\bibitem[Zhang et~al.(2022)Zhang, Qin, Zhang, Bao, Zhang, and Liu]{trans22}
Zhang, Q., Qin, C., Zhang, Y., Bao, F., Zhang, C., and Liu, P.
\newblock Transformer-based attention network for stock movement prediction.
\newblock \emph{Expert Syst. Appl.}, 202:\penalty0 117239, 2022.

\end{thebibliography}
\bibliographystyle{icml2025}

%%%%%%%%%%%%%%%%%%%%%%%%%%%%%%%%%%%%%%%%%%%%%%%%%%%%%%%%%%%%%%%%%%%%%%%%%%%%%%%
%%%%%%%%%%%%%%%%%%%%%%%%%%%%%%%%%%%%%%%%%%%%%%%%%%%%%%%%%%%%%%%%%%%%%%%%%%%%%%%
% APPENDIX
%%%%%%%%%%%%%%%%%%%%%%%%%%%%%%%%%%%%%%%%%%%%%%%%%%%%%%%%%%%%%%%%%%%%%%%%%%%%%%%
%%%%%%%%%%%%%%%%%%%%%%%%%%%%%%%%%%%%%%%%%%%%%%%%%%%%%%%%%%%%%%%%%%%%%%%%%%%%%%%
% \newpage
% \appendix
% \onecolumn

% You can have as much text here as you want. The main body must be at most $8$ pages long.
% For the final version, one more page can be added.
% If you want, you can use an appendix like this one.  

% The $\mathtt{\backslash onecolumn}$ command above can be kept in place if you prefer a one-column appendix, or can be removed if you prefer a two-column appendix.  Apart from this possible change, the style (font size, spacing, margins, page numbering, etc.) should be kept the same as the main body.
%%%%%%%%%%%%%%%%%%%%%%%%%%%%%%%%%%%%%%%%%%%%%%%%%%%%%%%%%%%%%%%%%%%%%%%%%%%%%%%
%%%%%%%%%%%%%%%%%%%%%%%%%%%%%%%%%%%%%%%%%%%%%%%%%%%%%%%%%%%%%%%%%%%%%%%%%%%%%%%

\end{document}